\pdfoutput=1

\documentclass[11pt,a4paper]{scrartcl}

\usepackage{CLICdp}
\usepackage{CLICdp_definitions}

\title{Hidden Valley searches at CLIC}
\clicdpconf{2018}{011}  
\date{\formatdate{29}{11}{2018}}
\addauthor{Marcin Kucharczyk\thanks{Marcin.Kucharczyk@cern.ch}}{\institute{1}}
\addinstitute{1}{The Henryk Niewodniczanski Institute of Nuclear Physics, Polish Academy of Sciences, Krakow, Poland}

\abstract{The sensitivity studies to observe long-lived particles predicted by a set of beyond the Standard Model theoretical models are reported. The analysis is based on a data sample of $e^{+} e^{-}$ collisions at $\sqrt{s} = 3$~TeV, simulated with the CLIC\_ILD detector model and corresponding to an integrated luminosity of 3~ab$^{-1}$. Long-lived particle decay products are subsequently combined to reconstruct the parent bosons employing secondary vertices displaced from the beam axis. The upper limits on the production cross section for the long-lived particle lifetimes from 1 to 300~ps, masses between 25 and 50~GeV/c$^{2}$, and a parent Higgs mass of 126~GeV/c$^{2}$ are determined.}

\titlecomment{Talk presented at the International Workshop on Future Linear Colliders (LCWS2018), Arlington, Texas, 22-26 October 2018. C18-10-22.}
\titlecomment{This work was carried out in the framework of the CLICdp Collaboration} 

\graphicspath{ {}{} }

\begin{document}

\titlepage

\newcommand{\latex}{\LaTeX\xspace}
\lstset{defaultdialect=[LaTeX]TeX}

\section{Introduction}
\label{sec:Intro}

Among many theoretical descriptions of new phenomena beyond the Standard Model (SM) there are a class of models predicting the existence of new massive Long-Lived Particles (LLP) with a measurable flight distance, providing displaced vertices (DV) which can be efficiently reconstructed by the tracking system of the CLIC detector~\cite{clicILD,CLICDet}. One class of such models, the so-called Hidden Valley, is a consequence of the string-theory, allowing for additional $hidden$ gauge sector which couples to SM particles only at very high energies~\cite{hidValley1,hidValley2}. The Hidden Valley particles are predicted in some of these models to have non-zero lifetime and to decay into $b\bar{b}$, having unobservable partners that could serve as dark matter objects. In the present report the analysis of the Standard Model Higgs boson decaying into two Hidden Valley particles $h^{0} \rightarrow \pi^0_v \pi^0_v$ is described. The Higgs boson is produced in WW-fusion, $e^{+} e^{-} \rightarrow h^{0} \nu_{e} \bar{\nu}_{e}$, at $\sqrt{s}$ = 3 TeV, simulated with the CLIC\_ILD detector model~\cite{CLICD}, using an integrated luminosity of 3~ab$^{-1}$. Similar searches of the SM Higgs boson decaying into two LLP's providing two $b\bar{b}$ di-jets in the final state have already been reported by the D0~\cite{hvD0}, CDF~\cite{hvCDF}, ATLAS~\cite{hvATLAS}, CMS~\cite{hvCMS} and LHCb~\cite{hvLHCb} experiments.

\section{Analysis strategy}
\label{sec:evtSel}

\begin{figure}[ht]
\begin{center}
\includegraphics[width=0.42\linewidth]{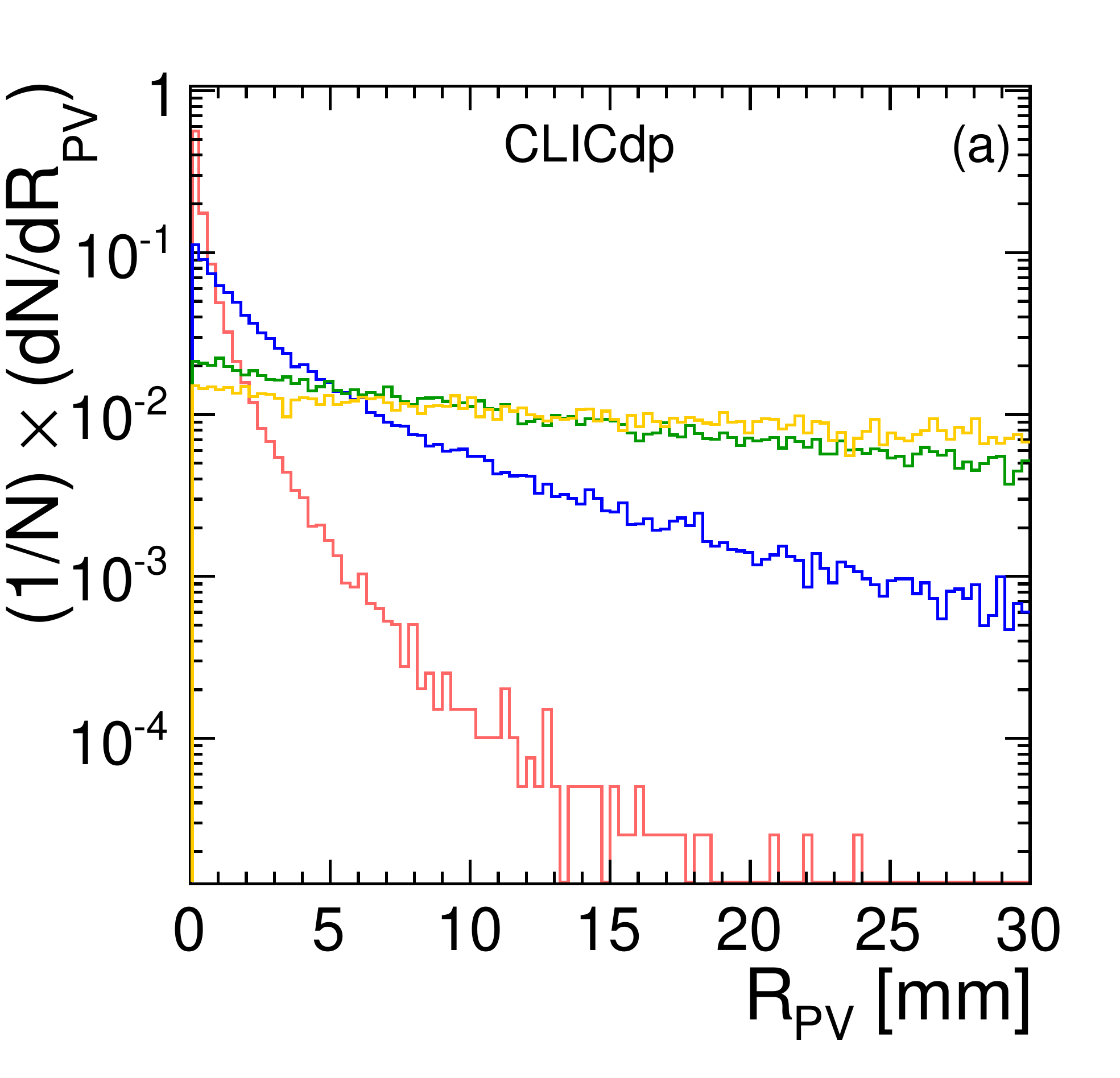}
\includegraphics[width=0.42\linewidth]{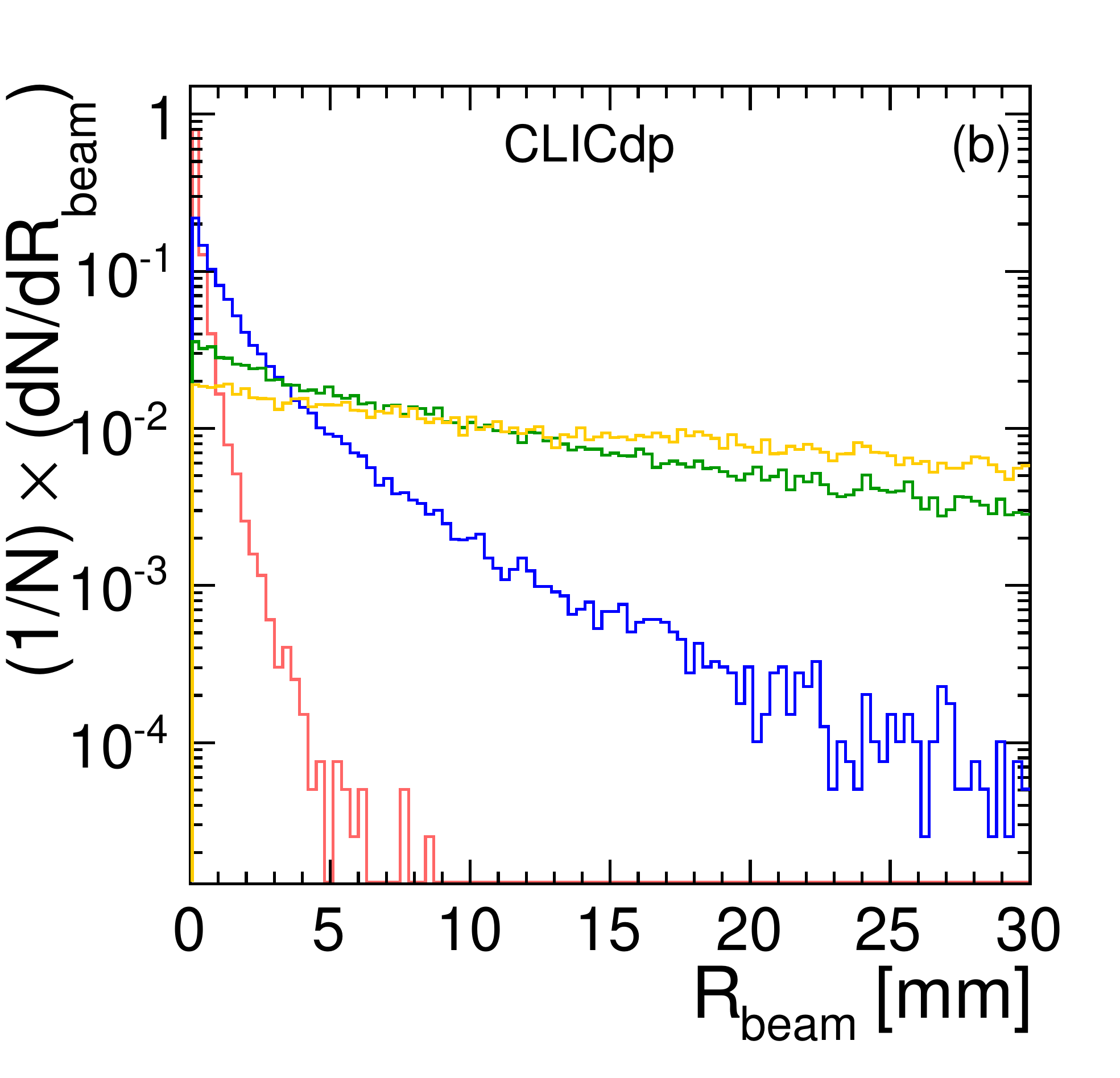}
\end{center}
\caption{Distance of the generated $\pi^0_v$ to (a) the PV and (b) its radial distance to the beam axis for $\pi^0_v$'s generated with a mass of 50 GeV/c$^{2}$ and with four different lifetimes: 1~ps (red line), 10~ps (blue line), 100~ps (green line) and 300~ps (yellow line). Figures adopted from~\cite{CLICHidValley}.}
\label{fig:genHVr}
\end{figure}

As the Hidden Valley objects are predicted to be massive and have non-zero lifetime, the possible presence of new long-lived particles is investigated based on events with reconstructed vertices displaced from the beam axis. In the process $h^{0} \rightarrow \pi^0_v \pi^0_v$, in order to reconstruct the parent Higgs boson, two $b$-tagged jets are assigned to each displaced vertex~\cite{CLICHidValley}. A sensitivity of the CLIC\_ILD detector to LLP production is studied using  Monte Carlo (MC) events generated with WHIZARD 1.95~\cite{Whizard} and PYTHIA 6~\cite{Pythia}, configured to produce Hidden Valley processes. The interaction of the generated particles with the CLIC\_ILD detector and its response are implemented using the Geant4~\cite{Geant4} simulation package and the MOKKA~\cite{Mokka} detector description toolkit. Finally, the MARLIN software package~\cite{Marlin} is used for event reconstruction, with the track reconstruction as in Ref.~\cite{clicTrack}. The signal and background samples of $e^{+} e^{-}$ collisions at $\sqrt{s}$ = 3 TeV are generated, namely $h^{0} \rightarrow \pi^0_v \pi^0_v$ with $\pi^0_v$ lifetimes from 1 to 300~ps, masses between 25 and 50 GeV/c$^{2}$, and parent Higgs mass of 126 GeV/c$^{2}$, as well as $e^{+} e^{-} \rightarrow q \bar{q}$, $e^{+} e^{-} \rightarrow q \bar{q} \nu \bar{\nu}$, $e^{+} e^{-} \rightarrow q \bar{q} q \bar{q}$ and $e^{+} e^{-} \rightarrow q \bar{q} q \bar{q} \nu \bar{\nu}$. For the signal samples a cross section of 0.42~pb is assumed, while for the background samples it is 2.95~pb, 1.32~pb, 0.55~pb and 0.07~pb for $e^{+} e^{-} \rightarrow q \bar{q}$, $e^{+} e^{-} \rightarrow q \bar{q} \nu \bar{\nu}$, $e^{+} e^{-} \rightarrow q \bar{q} q \bar{q}$ and $e^{+} e^{-} \rightarrow q \bar{q} q \bar{q} \nu \bar{\nu}$, respectively. The distance of the generated $\pi^0_v$ to the primary vertex (PV) and the radial distance to the beam axis are illustrated in Fig.~\ref{fig:genHVr}, showing the dependence on the $\pi^0_v$ lifetime.

\begin{figure}[h]                                                               
\begin{center}                                                                  
\includegraphics[width=0.42\linewidth]{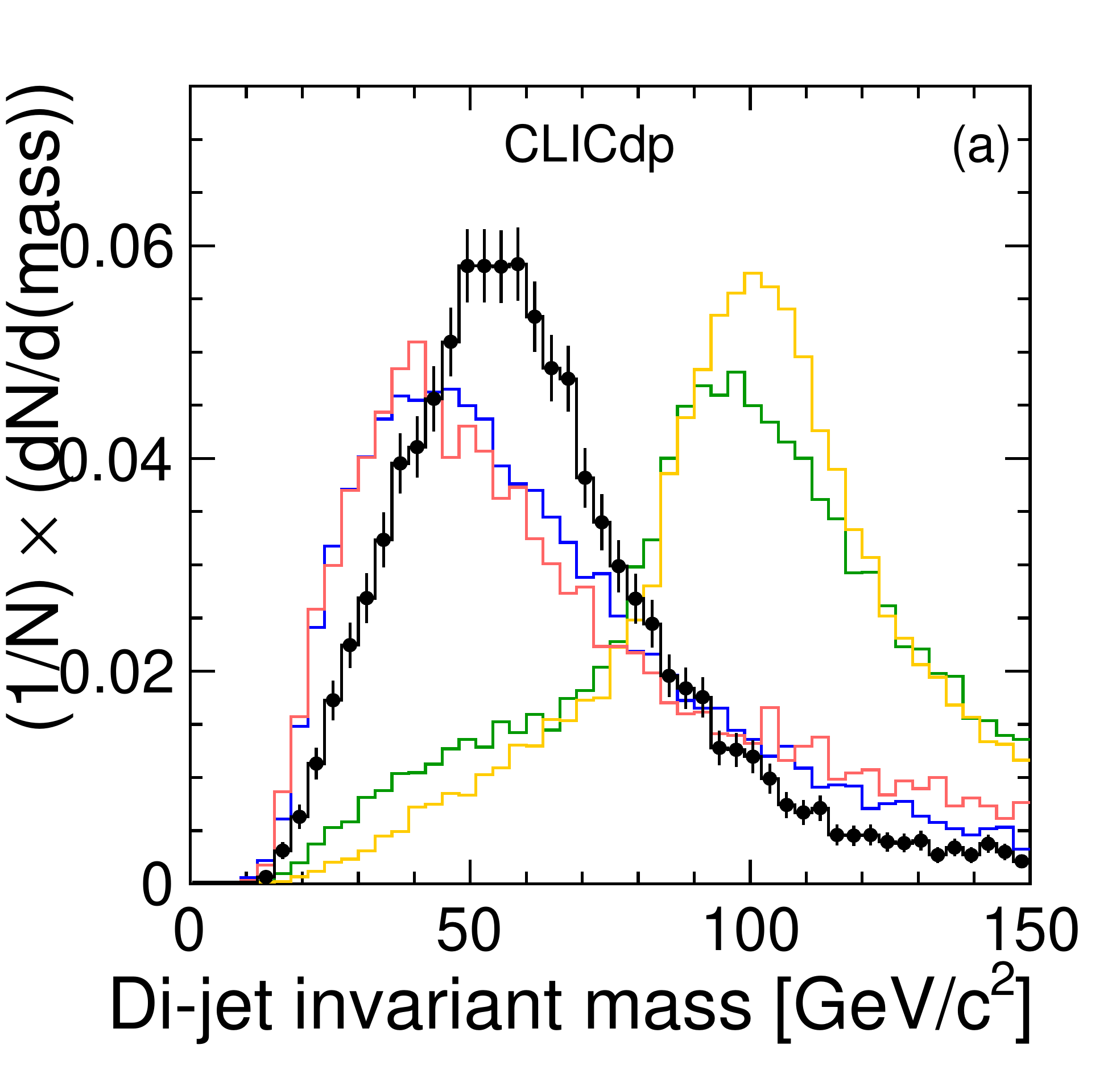}                               
\includegraphics[width=0.42\linewidth]{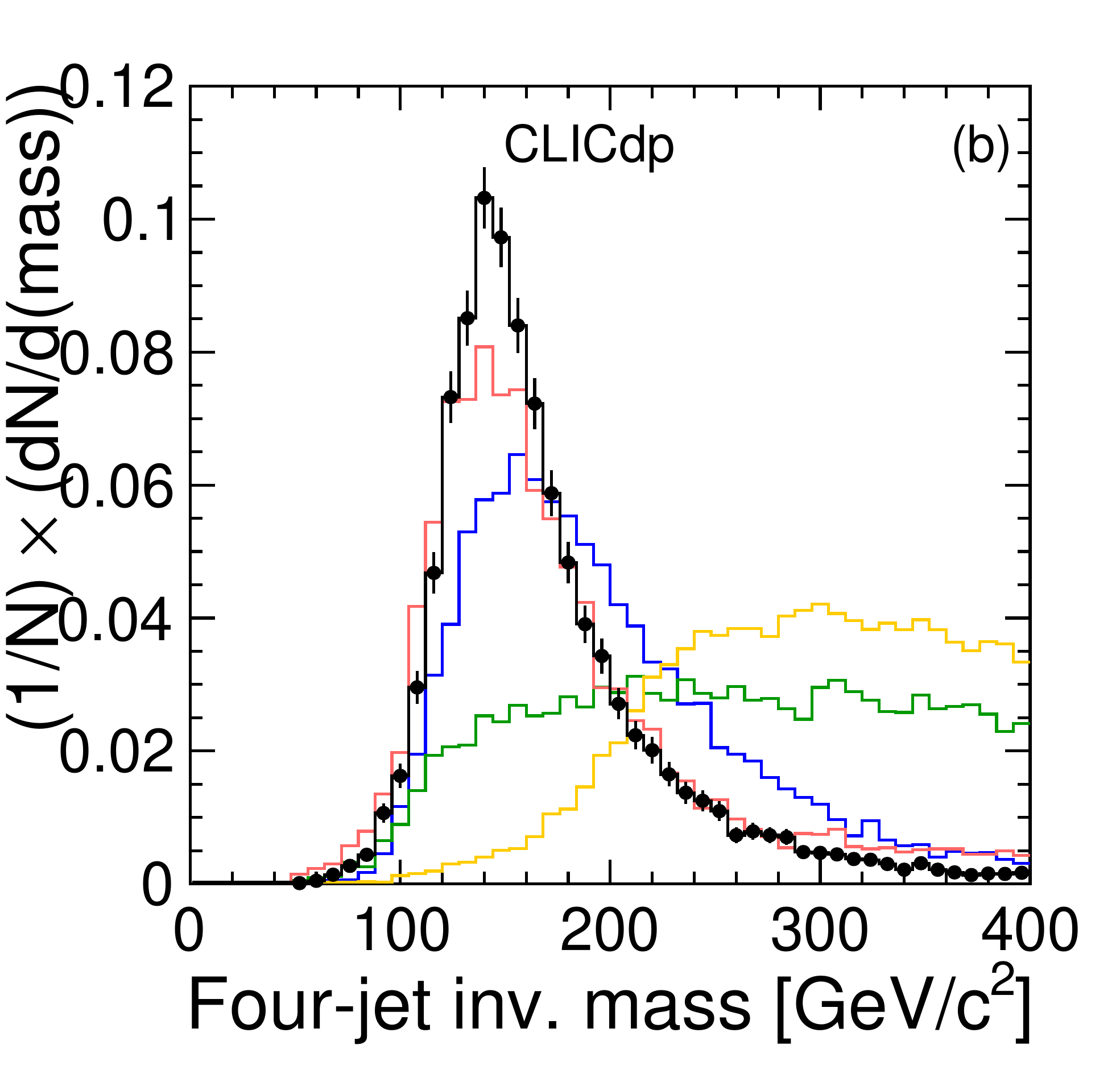}                             
\end{center}                                                                    
\caption{Di-jet (a) and four-jet (b) invariant mass for the signal sample of $\pi^0_v$ with a mass of 50~GeV/c$^{2}$ and with a lifetime of 10~ps (black) and the background of $q \bar{q}$ (red), $q \bar{q} \nu \bar{\nu}$ (blue), $q \bar{q} q \bar{q}$ (green) and $q \bar{q} q \bar{q} \nu \bar{\nu}$ (yellow). Figures adopted from~\cite{CLICHidValley}.}    
\label{fig:difourJetmass}                                                       
\end{figure}

The particles are reconstructed using default CLICdp packages~\cite{PFA}, while the jets are reconstructed using the longitudinally invariant $k_{t}$ algorithm~\cite{kt} as implemented in the FastJet package~\cite{Fastjet}. The jets are $b$- and $c$-tagged by passing a selection of parameters (including the impact parameters that have been computed via the vertexing) through a Boosted Decision Tree (BDT)~\cite{BDT}. The requirements on the particle impact parameter components as well as the value of the $R$ parameter used for the jet reconstruction are optimized according to the needs of the Hidden Valley analysis. For the Hidden Valley analysis it was also required to find all tracks from the $\pi^0_v$ decays, which are expected to be distant from the beam axis. A dedicated procedure to reconstruct displaced vertices has been developed and optimised for the Hidden Valley analysis (see~\cite{CLICHidValley} for details), as the default CLICdp vertexing~\cite{LCFI} could not efficiently find all tracks from the $\pi^0_v$ decays. The comparison of the di-jet and four-jet invariant mass distribution for the signal sample of $\pi^0_v$ with a mass of 50~GeV/c$^{2}$ and with a lifetime of 10~ps with the distributions for $q \bar{q}$, $q \bar{q} \nu \bar{\nu}$, $q \bar{q} q \bar{q}$ and $q \bar{q} q \bar{q} \nu \bar{\nu}$ backgrounds is shown in Fig.~\ref{fig:difourJetmass}, where the jets are reconstructed and $b$-tagged with $R = 1.0$. Di-jets are constructed by pairing two jets with an invariant mass closest to the $\pi^0_v$ mass.

Finally, two reconstructed jets which are positively tagged as $b$-jets are assigned to each displaced vertex. The di-jets are assigned to a single displaced vertex if the number of common charged tracks shared by this DV and the two $b$-tagged jets is maximum. The efficiencies for the signal and background samples with respect to the requirement to have at least two reconstructed displaced vertices in the event are listed in Table~\ref{tab:eff2DVs}.

\begin{table}[!htb]
\caption{The efficiencies of the requirement to have at least two reconstructed displaced vertices in the event for signal and background samples.}
\begin{center}\begin{tabular}{lccccc}
\hline
Process & $\pi^0_v$ lifetime [ps] & $\pi^0_v$ mass [GeV/c$^{2}$] & fraction of events with $\geq$ 2 DVs [\%]\\
\hline
$h^{0} \rightarrow \pi^0_v \pi^0_v$ & 1   & 25 & 68\\
$h^{0} \rightarrow \pi^0_v \pi^0_v$ & 10  & 25 & 86\\
$h^{0} \rightarrow \pi^0_v \pi^0_v$ & 100 & 25 & 93\\
$h^{0} \rightarrow \pi^0_v \pi^0_v$ & 300 & 25 & 80\\
$h^{0} \rightarrow \pi^0_v \pi^0_v$ & 1   & 35 & 70\\
$h^{0} \rightarrow \pi^0_v \pi^0_v$ & 10  & 35 & 86\\
$h^{0} \rightarrow \pi^0_v \pi^0_v$ & 100 & 35 & 94\\
$h^{0} \rightarrow \pi^0_v \pi^0_v$ & 300 & 35 & 82\\
$h^{0} \rightarrow \pi^0_v \pi^0_v$ & 1   & 50 & 72\\
$h^{0} \rightarrow \pi^0_v \pi^0_v$ & 10  & 50 & 89\\
$h^{0} \rightarrow \pi^0_v \pi^0_v$ & 100 & 50 & 90\\
$h^{0} \rightarrow \pi^0_v \pi^0_v$ & 300 & 50 & 86\\
\hline
$e^{+} e^{-} \rightarrow q \bar{q}$ & - & - & 6\\
$e^{+} e^{-} \rightarrow q \bar{q} \nu \bar{\nu}$ & - & - & 8\\
$e^{+} e^{-} \rightarrow q \bar{q} q \bar{q}$ & - & - & 9\\
$e^{+} e^{-} \rightarrow q \bar{q} q \bar{q} \nu \bar{\nu}$ & - & - & 11\\
\hline
\end{tabular}\end{center}
\label{tab:eff2DVs}
\end{table}

\section{Background reduction}
\label{sec:separation}

\begin{figure}[h]
\begin{center}
\includegraphics[width=0.42\linewidth]{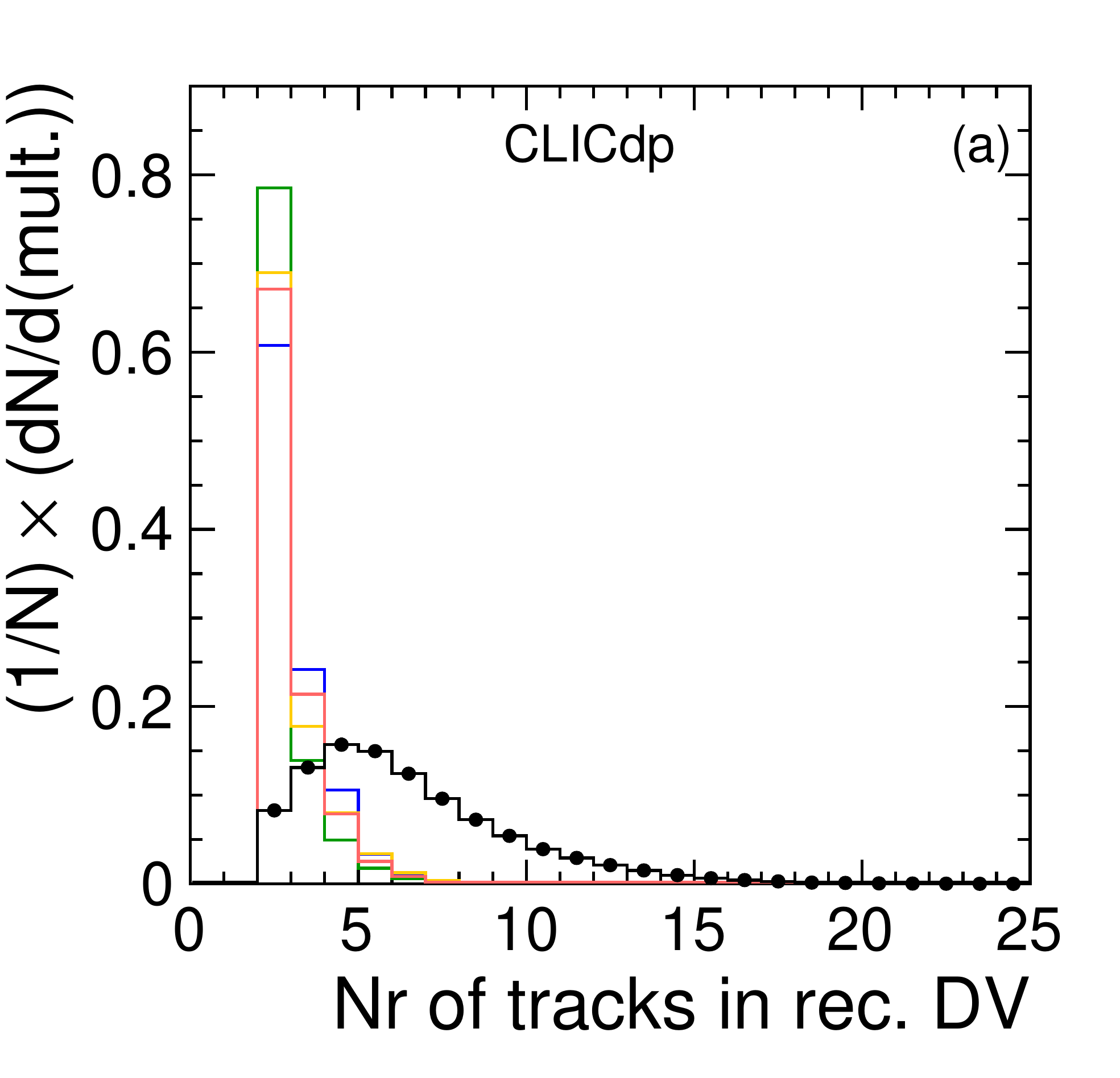}
\includegraphics[width=0.42\linewidth]{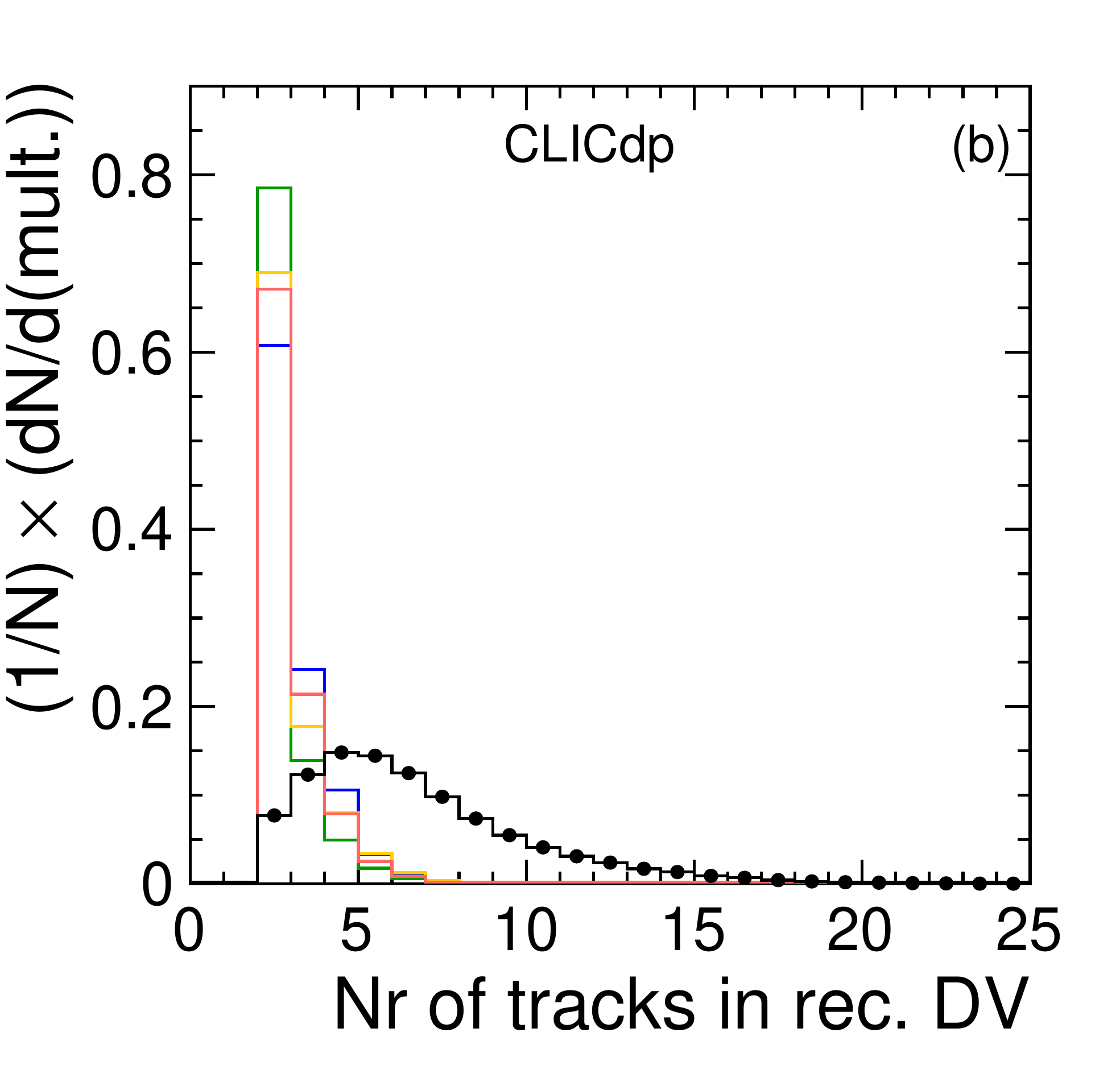}
\includegraphics[width=0.42\linewidth]{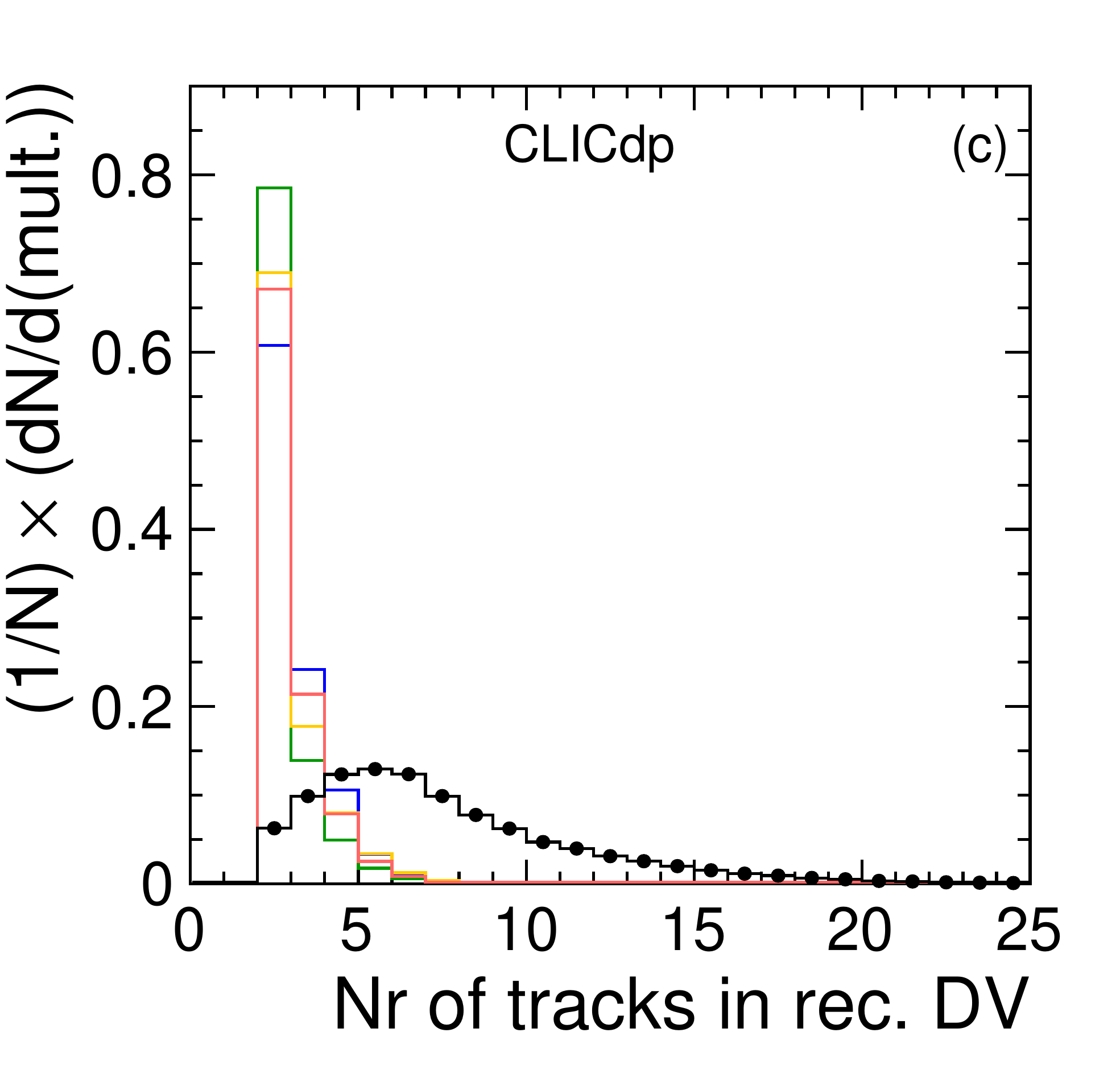}
\end{center}
\caption{Number of tracks assigned to the reconstructed DV for signal samples of $\pi^0_v$ with a mass of (a) 25~GeV/c$^{2}$, (b) 35~GeV/c$^{2}$ and (c) 50~GeV/c$^{2}$, and with a lifetime of 10~ps (black) compared to $q \bar{q}$ (red), $q \bar{q} \nu \bar{\nu}$ (blue), $q \bar{q} q \bar{q}$ (green) and $q \bar{q} q \bar{q} \nu \bar{\nu}$ (yellow) background events. Figures adopted from~\cite{CLICHidValley}.}
\label{fig:multTrDV}
\end{figure}

\begin{figure}[h]
\begin{center}
\includegraphics[width=0.42\linewidth]{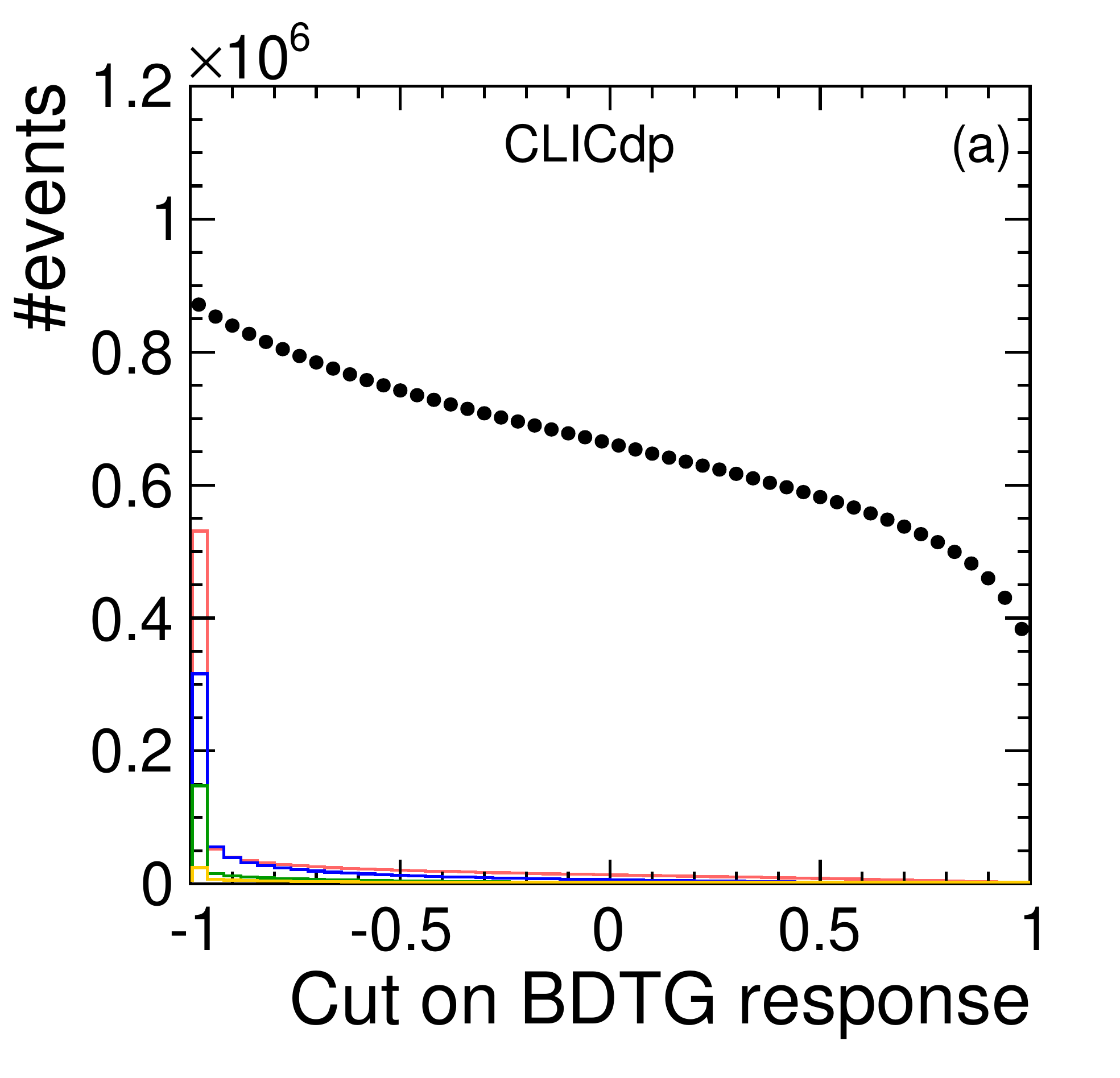}
\includegraphics[width=0.42\linewidth]{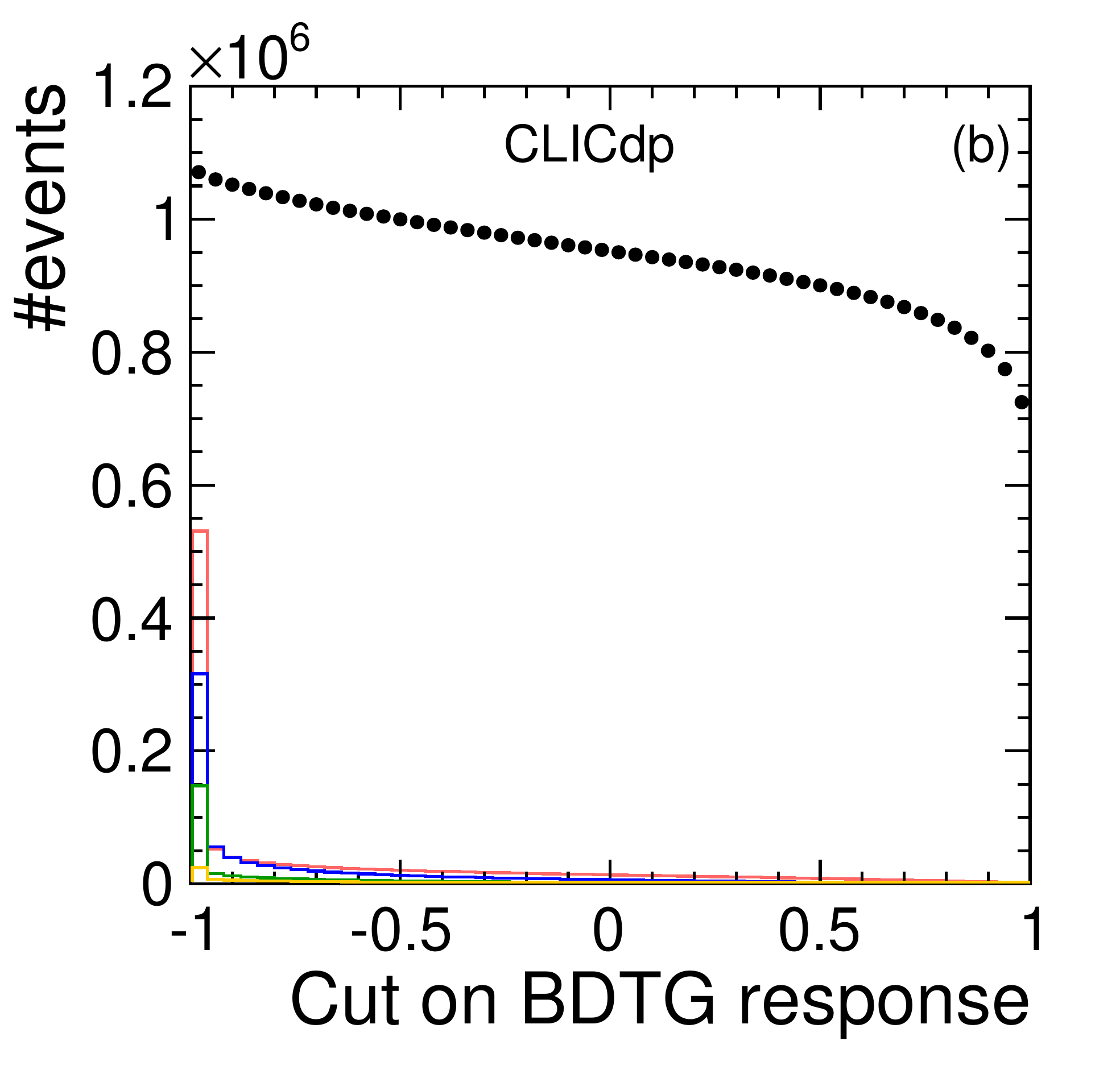}
\includegraphics[width=0.42\linewidth]{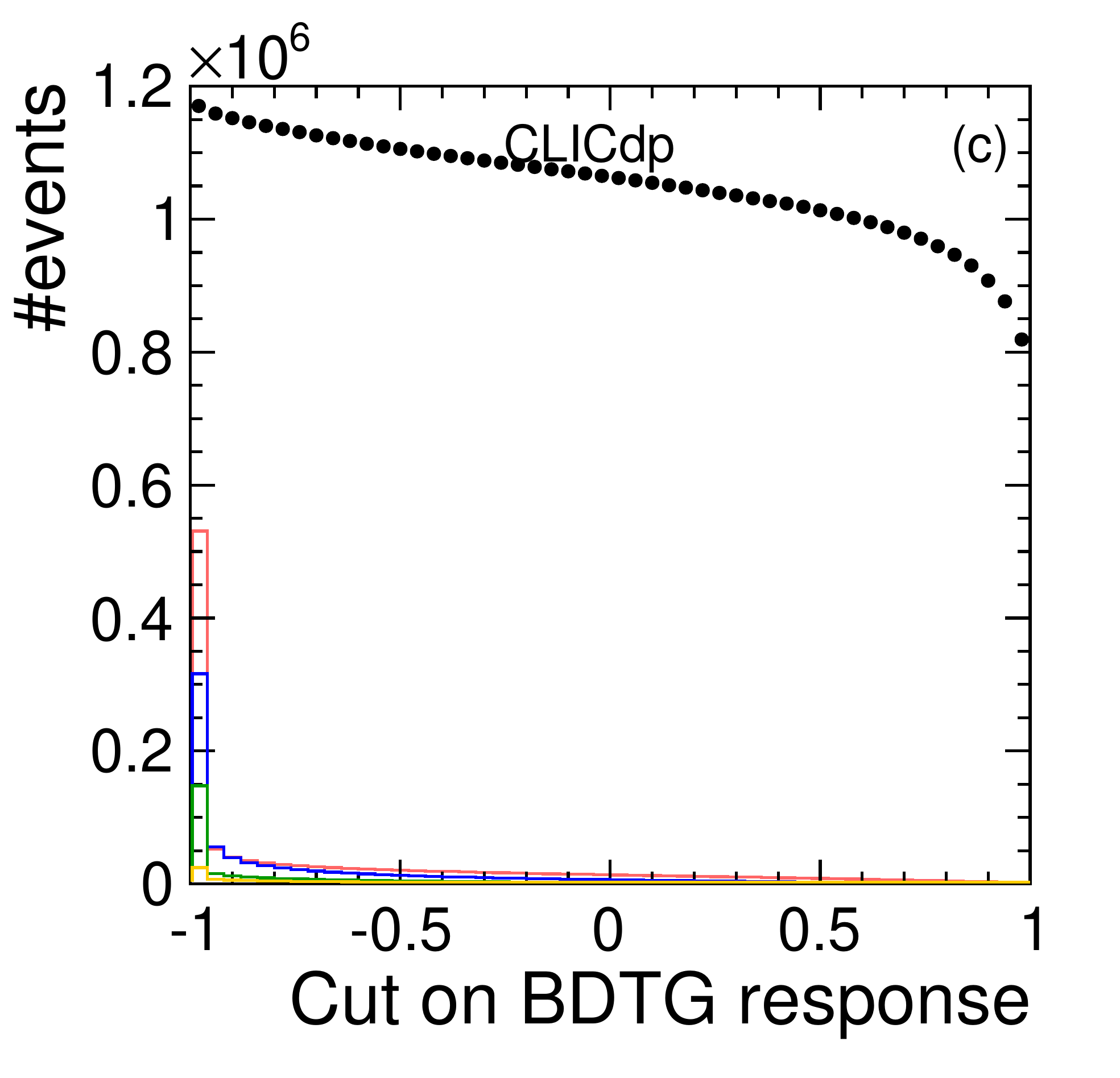}
\includegraphics[width=0.42\linewidth]{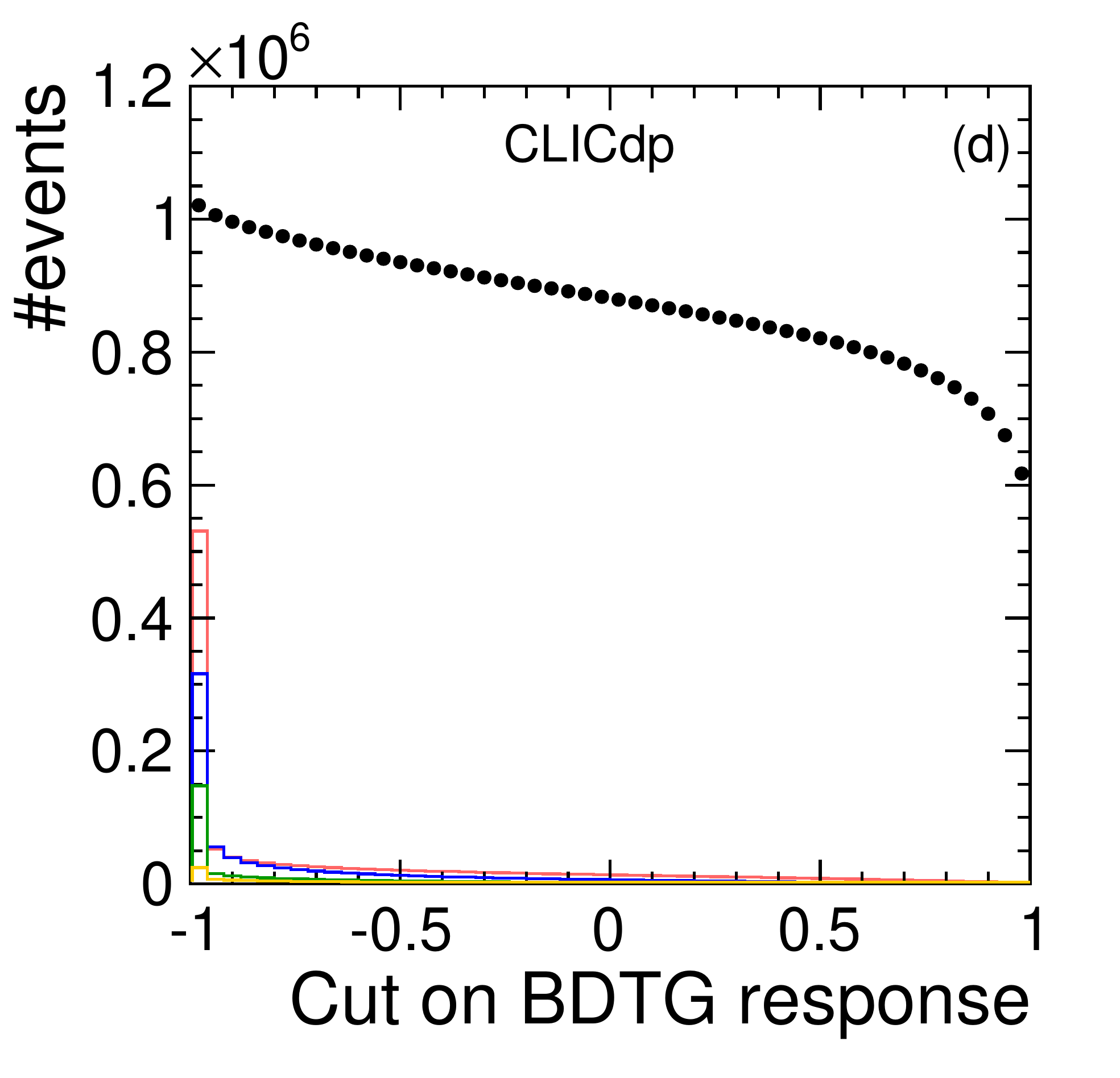}
\end{center}
\caption{Distributions of the expected number of events as a function of the cut on the BDTG response, for the assumed integrated luminosity of 3~ab$^{-1}$, for signal samples of $\pi^0_v$ with a mass of 35~GeV/c$^{2}$ (black figures) and four different lifetimes: (a) 1~ps, (b) 10~ps, (c) 100~ps, (d) 300~ps. Distributions of four different types of background are also plotted: $q \bar{q}$ (red), $q \bar{q} \nu \bar{\nu}$ (blue), $q \bar{q} q \bar{q}$ (green) and $q \bar{q} q \bar{q} \nu \bar{\nu}$ (yellow). Figures adopted from~\cite{CLICHidValley}.}
\label{fig:BDTGcut35}
\end{figure}

In order to separate $\pi^0_v$ signal from the $q \bar{q}$, $q \bar{q} \nu \bar{\nu}$, $q \bar{q} q \bar{q}$, $q \bar{q} q \bar{q} \nu \bar{\nu}$ background a multivariate analysis based on the Boosted Decision Tree Gradient (BDTG)~\cite{TMVA} is applied. It employs seven variables providing significant signal to background separation, i.e. the number of tracks assigned to the reconstructed DV, the number of reconstructed DVs in the event, the invariant mass of the DV, the di-jet invariant mass of two jets assigned to the DV, the four-jet invariant mass of two di-jets assigned to two DVs, the distance $y_{n+1,n}$ at which the transition from a three jet event to a two jet event takes place, the distance $y_{n-1,n}$ at which the transition from a four jet event to a three jet event takes place. Fig.~\ref{fig:multTrDV} shows as example the number of tracks assigned to the reconstructed DV for the signal sample of $\pi^0_v$ with three different masses and the lifetime of 10~ps, and for the background of $q \bar{q}$, $q \bar{q} \nu \bar{\nu}$, $q \bar{q} q \bar{q}$ and $q \bar{q} q \bar{q} \nu \bar{\nu}$ events. The Boosted Decision Tree Gradient method has been chosen as the most effective, where $\pi^0_v$ with a mass of 50~GeV/c$^{2}$ and lifetime of 10~ps has been used as a signal, and $e^{+} e^{-} \rightarrow q \bar{q} \nu \bar{\nu}$ as a background. Distributions of the expected number of events as a function of the cut on the BDTG response, for the assumed integrated luminosity of 3~ab$^{-1}$, with the cross sections mentioned in Sec.~\ref{sec:evtSel} and selection efficiencies taken from Table~\ref{tab:eff2DVs}, are shown in Fig.~\ref{fig:BDTGcut35} for $\pi^0_v$ mass hypotheses of 35~GeV/c$^{2}$ and four different lifetimes.

\section{Sensitivity and upper limits}
\label{sec:sensitivity}

\begin{figure}[h]
\begin{center}
\includegraphics[width=0.42\linewidth]{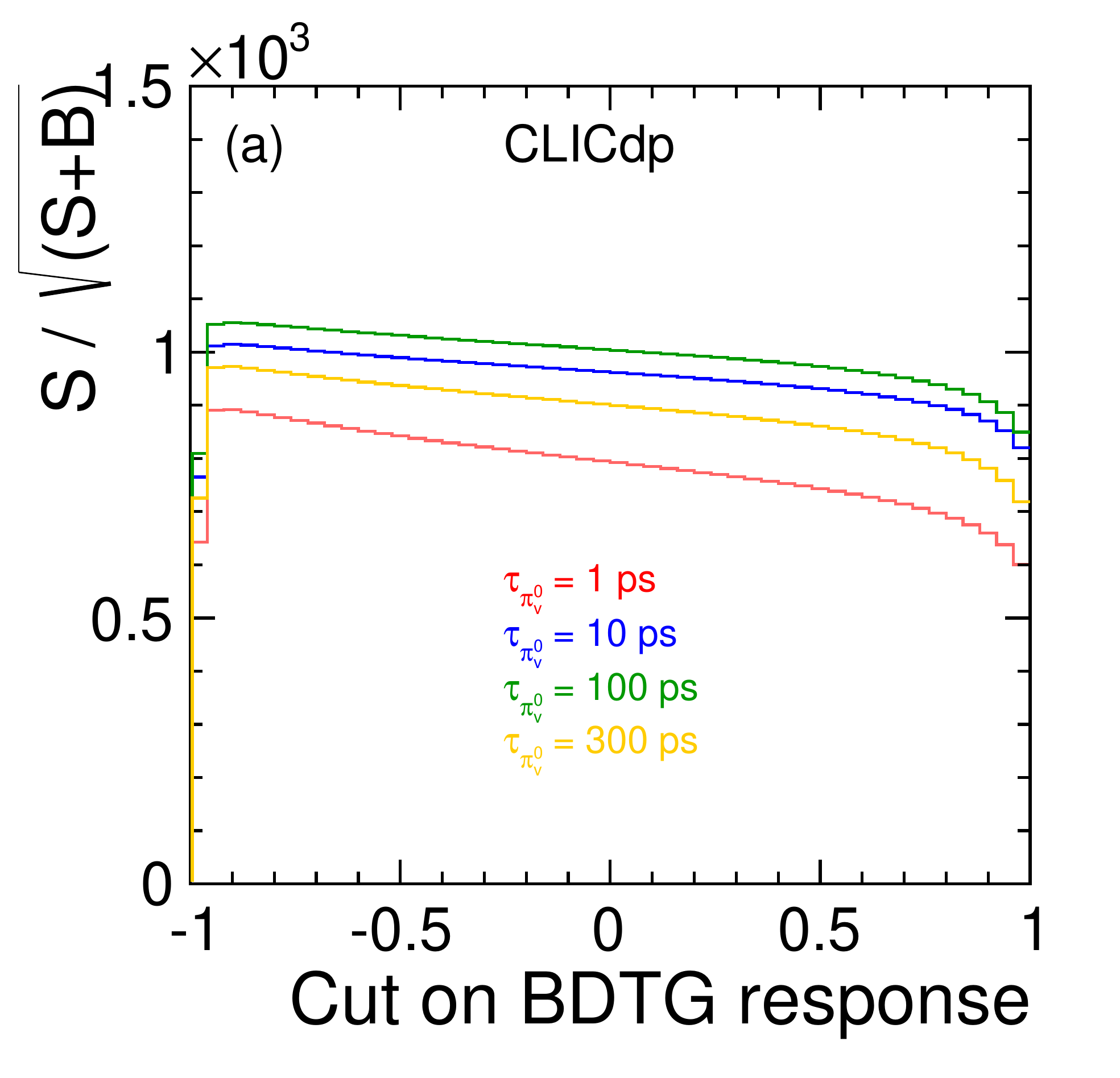}
\includegraphics[width=0.42\linewidth]{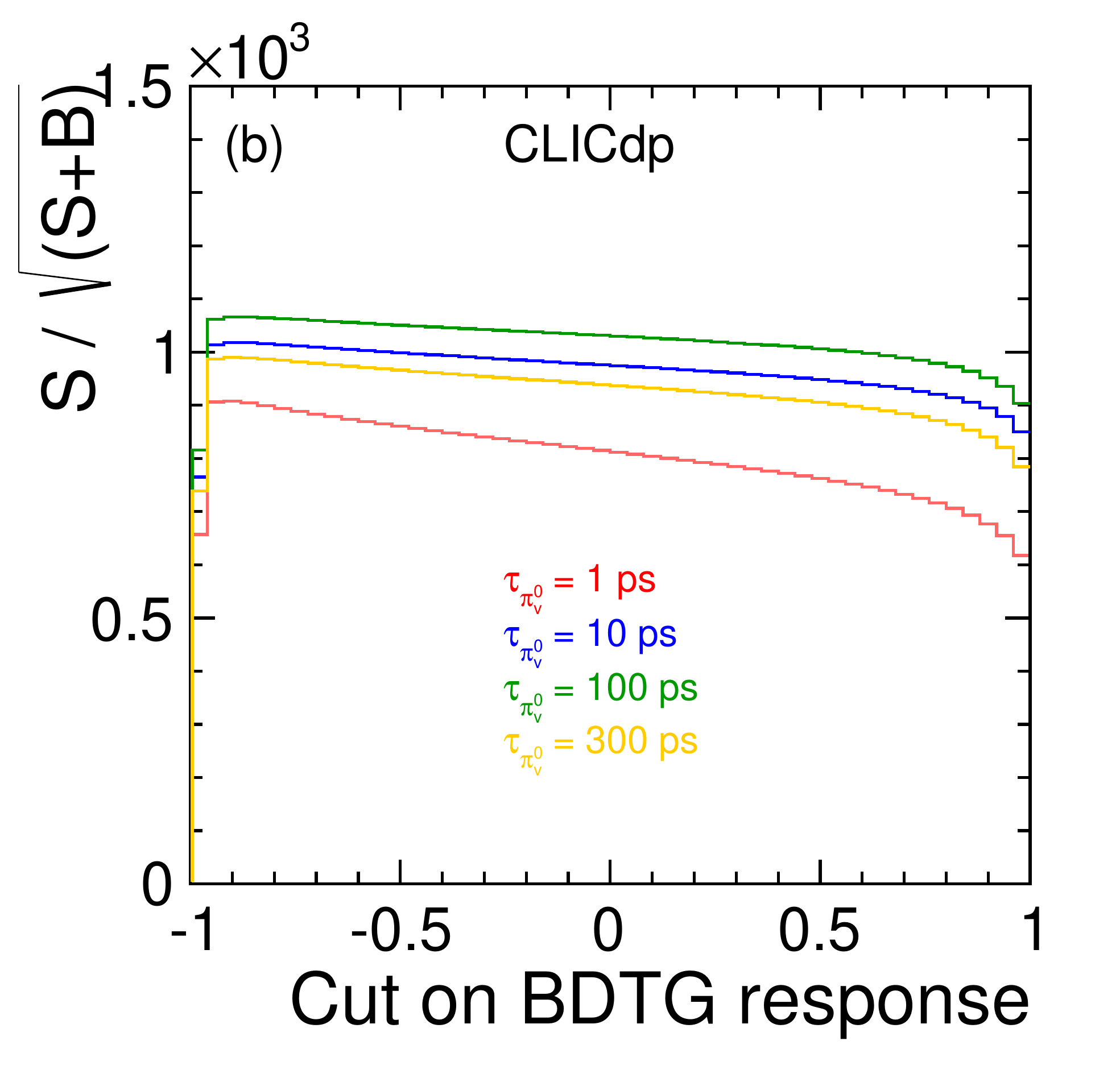}
\includegraphics[width=0.42\linewidth]{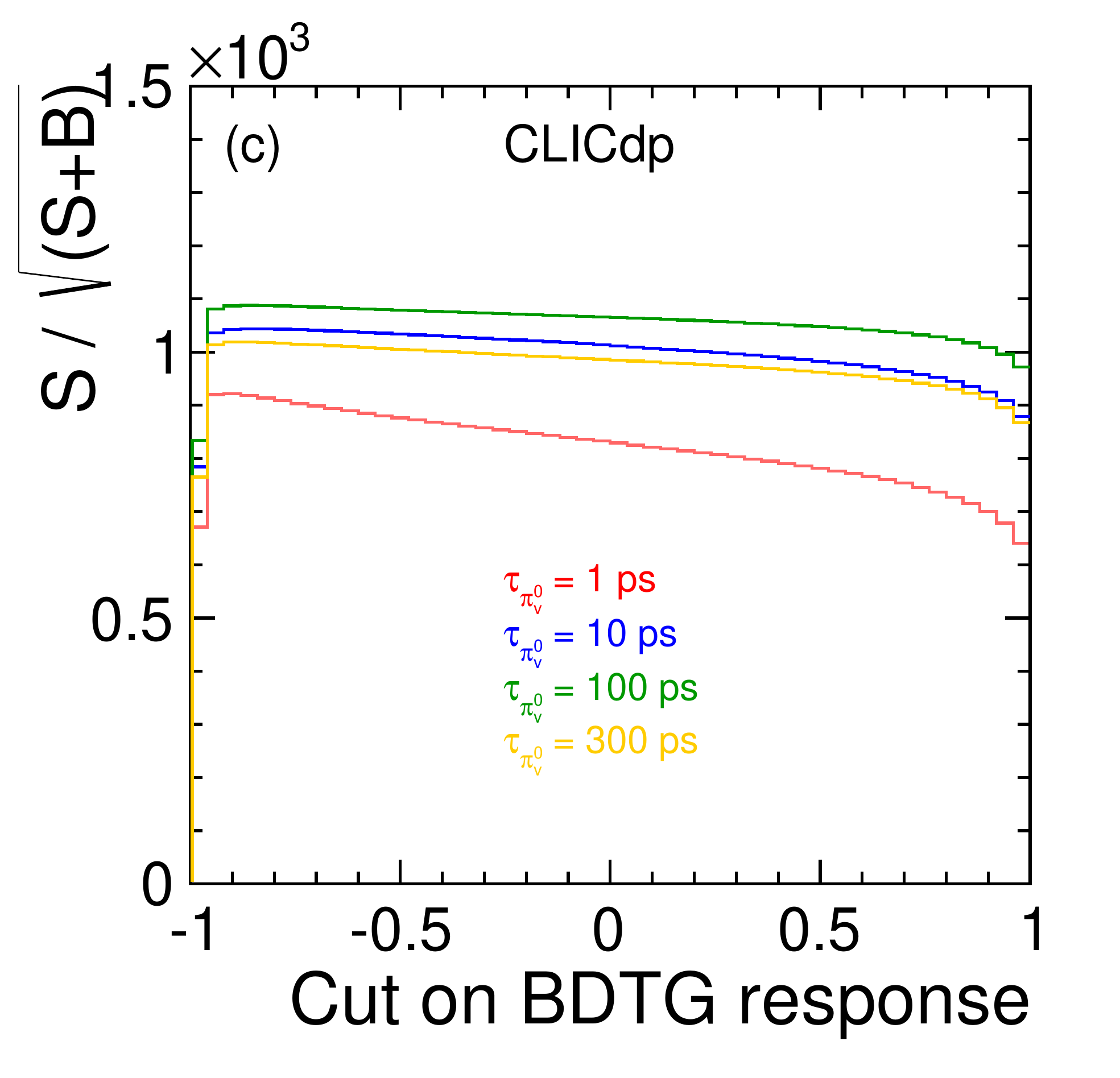}
\end{center}
\caption{Sensitivity $S / \sqrt{S+B}$ for the expected number of events as a function of the cut on the BDTG response, for signal samples of $\pi^0_v$ with a mass of (a) 25~GeV/c$^{2}$, (b) 35~GeV/c$^{2}$ and (c) 50~GeV/c$^{2}$, and for four different lifetimes: 1~ps (red line), 10~ps (blue line), 100~ps (green line) and 300~ps (yellow line). Figures adopted from~\cite{CLICHidValley}.}
\label{fig:sensitivity}
\end{figure}

For all the $\pi^0_v$ mass and lifetime hypotheses the sensitivity of the CLIC\_ILD detector to observe $\pi^0_v$ particles through the Higgs boson decay $h^{0} \rightarrow \pi^0_v \pi^0_v$ has been estimated. The Higgs boson production via WW-fusion process has been assumed, $e^{+} e^{-} \rightarrow h^{0} \nu_{e} \bar{\nu}_{e}$, at $\sqrt{s}$ = 3 TeV and integrated luminosity of 3~ab$^{-1}$. Fig.~\ref{fig:sensitivity} shows the sensitivity as a function of the cut on the BDTG response for signal samples with different masses andlifetimes, where four different sources of background are added: $e^{+} e^{-} \rightarrow q \bar{q}$, $e^{+} e^{-} \rightarrow q \bar{q} \nu \bar{\nu}$, $e^{+} e^{-} \rightarrow q \bar{q} q \bar{q}$, $e^{+} e^{-} \rightarrow q \bar{q} q \bar{q} \nu \bar{\nu}$. The requirement on BDTG response to be $>$~0.95 for all the $\pi^0_v$ mass and lifetime configurations is taken in order to almost completely remove the backgrounds, keeping nearly the entire signal.

\begin{figure}[h]
\begin{center}
\includegraphics[width=0.42\linewidth]{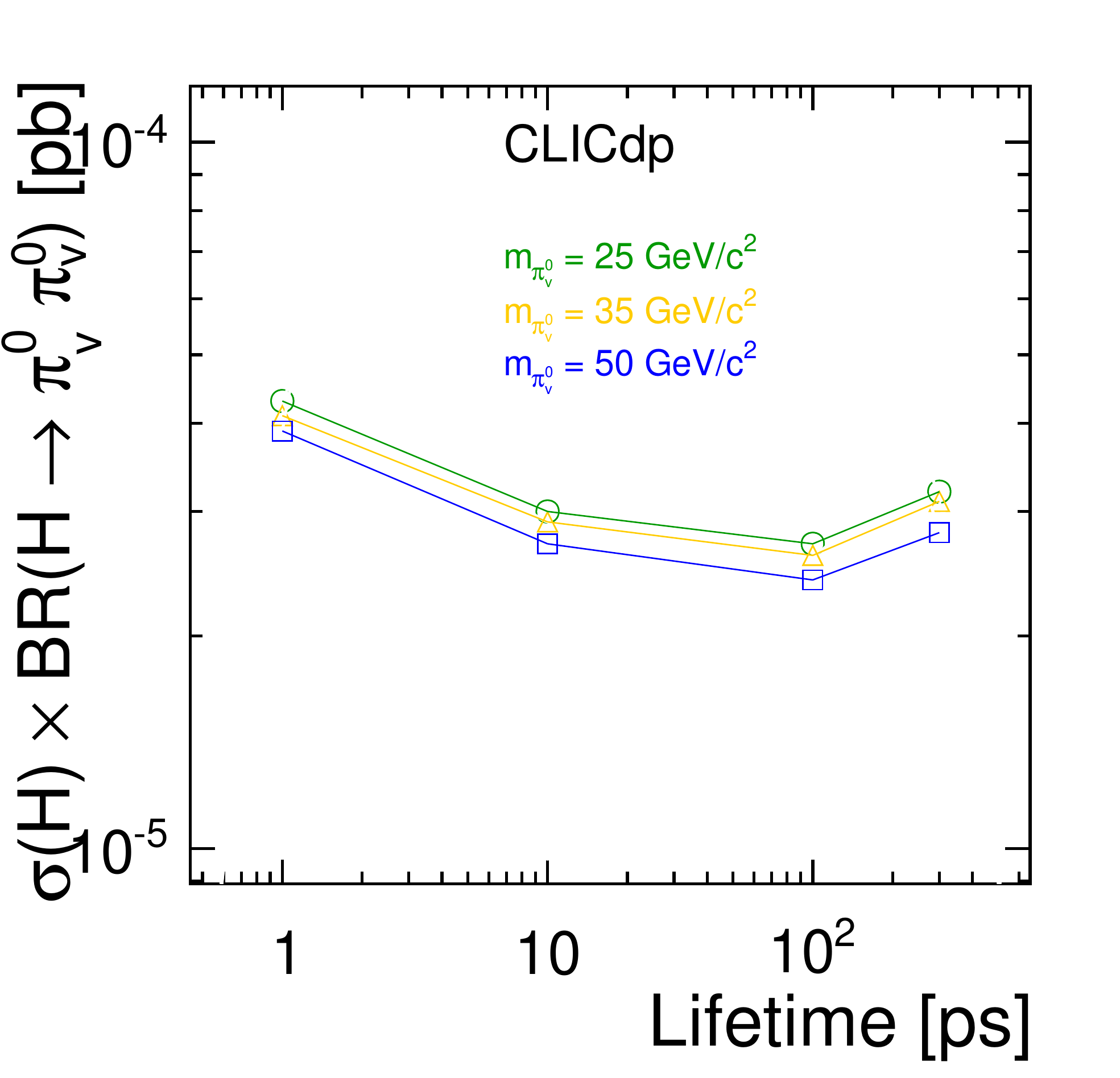}
\end{center}
\caption{Observed 95\% CL cross-section upper limits on the $\sigma(H) \times BR(H \rightarrow \pi^0_v \pi^0_v)$, within the model~\cite{hidValley2}, for three different $\pi^0_v$ masses: 25~GeV/c$^{2}$ (green), 35~GeV/c$^{2}$ (yellow), 50~GeV/c$^{2}$ (blue), as a function of $\pi^0_v$ lifetime. Figure adopted from~\cite{CLICHidValley}.}
\label{fig:upl}
\end{figure}

Finally, the upper limits on the product of the Higgs production cross-section and the branching fraction of the Higgs boson decay into long-lived particles, $\sigma(H) \times BR(H \rightarrow \pi^0_v \pi^0_v)$ have been determined employing the CL(s) method~\cite{CLs}. Under the assumption of a 100\% branching fraction for $\pi^0_v \rightarrow b\bar{b}$ and the cut on the BDTG response $> 0$, the 95\% CL upper limits on $\sigma(H) \times BR(H \rightarrow \pi^0_v \pi^0_v)$ have been computed. Fig.~\ref{fig:upl} shows the observed 95\% CL cross-section upper limits on the $\sigma(H) \times BR(H \rightarrow \pi^0_v \pi^0_v)$, within the model~\cite{hidValley2}, for three different masses, as a function of $\pi^0_v$ lifetime.

\section{Conclusions}
\label{sec:separation}

In the present report results of the search for the exotic LLPs within the CLIC\_ILD detector are described, for the SM Higgs boson decaying to a pair of LLPs, using $e^{+} e^{-}$ event sample simulated at $\sqrt{s}$ = 3 TeV and corresponding to an integrated luminosity of 3~ab$^{-1}$. The analysis based on reconstructed displaced vertices provides an immense reduction of the large Standard Model background using a multivariate analysis approach. The upper limits obtained are much more precise compared to those of the currently operating detectors~\cite{hvATLAS,hvCMS,hvLHCb}.

\section{Acknowledgements}
This work benefited from services provided by the ILC Virtual Organisation, supported by the national resource providers of the EGI Federation. This research was done using resources provided by the Open Science Grid, which is supported by the National Science Foundation and the U.S.~Department of Energy's Office of Science.

\printbibliography[title=References]

\end{document}